 \documentclass[final,3p,times]{elsarticle}

\def\<{\langle}
\def\>{\rangle}
\def\all{\allowdisplaybreaks}

\def\RR{\mathbb{R}}
\def\NN{\mathbb{N}}

\def\LL{\mathbb{L}}

\def\cas#1{\begin{cases}#1\end{cases}}

\def\bal#1\eal{\all\begin{align}#1\end{align}}
\def\bals#1\eals{\all\begin{align*}#1\end{align*}}
\def\bald#1\eald{\all\begin{aligned}#1\end{aligned}}
\def\barr#1\earr{\begin{array}#1\end{array}}
\def\be#1\ee{\begin{equation}#1\end{equation}}
\usepackage{amssymb}
\usepackage{amsmath} 
\newtheorem{theorem}{Theorem}

\begin{document}

\begin{frontmatter}



\title{An orthogonal basis expansion method for  solving path-independent stochastic differential equations }

\author[label1]{Rahman Farnoosh}
\ead{rfarnoosh@iust.ac.ir}
\address[label1]{School of Mathematics, Iran University of Science and Technology, 16844 Tehran,Iran.}
\author[label1]{Amirhossein Sobhani}
\ead{a\_sobhani@aut.ac.ir, a\_sobhani@mathdep.iust.ac.ir}
\author[label1]{Hamidreza Rezazadeh}
\ead{hr\_rezazadeh@MathDep.iust.ac.ir}

\begin{abstract}

In this article, we present an orthogonal basis expansion method for solving stochastic differential equations with a path-independent solution of the form $X_{t}=\phi(t,W_{t})$. For this purpose, we define a Hilbert space and construct an orthogonal basis for this inner product space with the aid of 2D-Hermite polynomials. With considering $X_{t}$ as orthogonal basis expansion, this method is implemented and the expansion coefficients are obtained by solving a system of nonlinear integro-differential equations. The strength of such a method is that expectation and variance of the solution is computed by these coefficients directly. Eventually, numerical results demonstrate its validity and efficiency in comparison with other numerical methods.
\end{abstract}


\end{frontmatter}


\section{Introduction}
\label{intro}
A stochastic  process $X_{t}$ on a filtered probability space $(\Omega, \mathcal{F} , \mathbb{P})$ is an It$\hat{o}$ process if it satisfies the following stochastic differential equation (SDE):
\bal\label{SDE}
\cas{dX_{t}=F(X_{t},t)dt+G(X_{t},t)dW_{t},\\X(0)=X_0. }
\eal
where $W_{t}$ is a standard Wiener process, the drift function $F: \RR^{}\times\RR^{+}\longrightarrow \RR^{}$ and the volatility function $G: \RR^{}\times\RR^{+}\longrightarrow \RR^{}$ are Borel measurable and locally bounded. In addition, we assume $F$ and $G$ are under linear growth and locally Lipschits conditions:
\bals
|F(X,t)-F(Y,t)|+|G(X,t)-G(Y,t)|\leq D|X-Y|~~~~,X,Y\in\RR^{}~~ and~~ t\in[0,T]
\eals
\bals
|F(X,t)|+|G(X,t)|\leq C(1+|X|)
\eals
these conditions imply equation \eqref{SDE} has a unique t-continuous solution adapted to filtration $\{\mathcal{F}_{t}\}_{t\geq0}$ generated by the Wiener process $W(t)$ ( see \cite{Oksendal}, \cite{Arnold}, \cite{platen}) and 
\bal
 E[\int_{0}^{T}|X(s)|^2ds]<\infty.
\eal
Also for simplicity $X_{0}$ is considered a non-random real number.
\par
   In the general case It$\hat{o}$ process $X_t$ depends on the history of Wiener path, $W_ {[0, t]} $ where $W_{[0,t]}=\lbrace W_{s}:0\leq s\leq t \rbrace$, in the special case that $X_ {t} $ depends only on $W_t$ (i.e,$X_{t}=\phi(t,W_{t})$), it is called a path-independent solution of equation \eqref{SDE}. Under some additional conditions for drift and volatility functions, if we have
   
   \begin{equation}
   \label{pa11}
   G\frac{\partial F}{\partial X}-F\frac{\partial G}{\partial X}-\frac{\partial G}{\partial t}-
  \frac{G^2}{2} \frac{\partial^2 G}{\partial X^2}=0
   \end{equation}
  equation\eqref{SDE} has a unique path-independent solution \cite{path}.
  \par
  Stochastic differential equations (SDEs) play a key role in modeling of phenomena that arise in vast variety of application areas including including Finance \cite{f}, Biology \cite{Allen}, Chemistry \cite{Allen1} and Physics\cite{Allen2}. Due to the fact that finding the analytical solution of SDEs is not easy, except for some special cases, the numerical simulation methods has become a favorite topic in the study of SDEs and devoted a lot of attention of researchers in recent years. At the present time several numerical methods such as \emph{Euler�Maruyama} (E.M.) , \emph{Milstein} \cite{T}, chain Rule \cite{Higam} and also stochastic Runge-Kutta methods\cite{B} has been proposed.
  \par
  In this paper, we propose an orthogonal basis expansion method for solving SDEs with path-independent solutions. The orthogonal basis expansion method is one of the most popular topics in functional analysis and widely implemented for solving Partial Differential Equations, Ordinary Differential Equations, and Integral Equations.
  \par
  For this purpose, we define a Hilbert space appropriately and assume that path-independent solution of equation\eqref{SDE} belongs to this space. Then we construct an orthogonal basis For this inner product space. Afterward, we consider $X_{t}$ in the form of orthogonal basis expansion. Finally, for finding the coefficients of this expansion, we reduce to a system of nonlinear integro-differential equations. The properties of this represented method
are similar to Wiener Chaos Expansion method (WCE) that is also referred to Hermite polynomial chaos expansion. (see, e.g. \cite{C}, \cite{J}, \cite{W} and the references therein). There is a fundamental different between these two methods in basis construction. Introduced basis in WCE method is multi-variable Hermite polynomials of the Gaussian random process that have been generated by tensor product but in this new method, we represent 2-dimensional Hermite polynomials.
\par
This paper is organized as follows. In section 2, we outline the theoretical foundation of stochastic Hermite polynomials as a basis and their properties. Section 3 constructs orthogonal basis expansion for functions of Wiener stochastic process and describes the general procedure by applying orthogonal basis expansion method. In addition, we find basic stochastic indicators of the solution such as expectation and variance. In section 4, numerical solution of some particular and well-known types of SDEs based on orthogonal basis expansion method is considered. Examples are surveyed and their exact solutions with both these solutions and other stochastic numerical simulation like predictor-corrector $ \emph{Euler�Maruyama}$ (E.M.) and $\emph{Milstein}'$ method are compared. Finally, a brief conclusion is stated in Section 5.
 \section{ Introducing a Stochastic Orthogonal Basis}
\label{sec:1}
 Consider the space of square integrable functions with respect to the time-dependent Gaussian weight function $\mu_{t}$:
 \bal
 \label{L2space}
 \LL ^{2}_{\gamma_t}(\RR\times[0,T])=\{ f(x,t);\int_{0}^{T}\int_{-\infty}^{\infty}f(x,t)^2\gamma_{t}dx dt<\infty,\}
 \eal
where 
\bal\label{measure}
\gamma_{t}= \frac{1}{\sqrt{2\pi t}}e^{\frac{-x^2}{2t}}.
\eal
  Furthermore, the inner product and norm of this Hilbert space are defined as usual:
\be
\<f,g\>_{\gamma_{t}}=\int_{0}^{T}\int_{-\infty}^{\infty}f(x,t)g(x,t)\gamma_{t}dx dt 
\ee
 \be
\| f \|_{\gamma_{t}}^{2}=\<f,f\>_{\gamma_{t}}=\int_{0}^{T}\int_{-\infty}^{\infty}f^2(x,t)\gamma_{t}dx dt 
\ee
Consider the Hermite polynomials
 \bal\label{h}
H_n(x)&=\frac{(-1)^n}{\sqrt{n!}}e^\frac{x^2}{2}\frac{d^n}{dx^n}(e^{\frac{-x^2}{2}})~,~( n = 0, 1,\cdots).
\eal
  The $n$-th order monic Hermite polynomials are reached by substituting $u=\dfrac{x}{\sqrt{t}}$ in \eqref{h}:
\bal\label{hermite}
\bar{H}_{n}(x,t) =(-t)^ne^\frac{x^2}{2t}\dfrac{d^n}{dx^n}(e^\frac{-x^2}{2t})=t^{\frac{n}{2}}H_{n}(\frac{x}{\sqrt{t}}).
\eal
 In $2-$dimensional, $n$-th order Hermite polynomials (\cite{Evance} and \cite{S}) are defined as follows:
\bal\label{hermite}
\mathcal{H}_{n}(x,t) =\dfrac{(-t)^n}{n!}e^\frac{x^2}{2t}\dfrac{d^n}{dx^n}(e^\frac{-x^2}{2t}),
\eal
 where $\mathcal{H}_{n}\in\LL^{2}(\RR\times(0,T] ,\rho)$ and it can be observed that,
 $$ \mathcal{H}_{n}(x,t)=\frac{1}{n!}\bar{H}_{n}(x,t)=\frac{\sqrt{t^n}}{n!}.H_{n}(\frac{x}{\sqrt{t}}).$$

 The \emph{monic Hermite polynomials} $\bar{H}_{n}(x,t)$ are generated by a three-term recursion relationship:

\begin{eqnarray}\label{recursive1}
\left\{ {\begin{array}{ll}\bar{H}_{n+1}(x,t)-x \bar{H}_{n}(x,t)+t n \bar{H}_{n-1}(x,t)=0~~,
\\\\ \ ~\bar{H}_{0}(x,t)=1~~~,~~~\bar{H}_{-1}(x,t)=0.
\end{array} }\right.
\end{eqnarray}
 From \eqref{recursive1}, we can get a similar recursion relationship of the $2-$ dimensional Hermite polynomials:
\begin{eqnarray}\label{recursive2}
\left\{ {\begin{array}{ll}(n+1) \mathcal{H}_{n+1}(x,t)-x \mathcal{H}_{n}(x,t)+t \mathcal{H}_{n-1}(x,t)=0~~,
\\\\ \ ~\mathcal{H}_{0}(x,t)=1~~~,~~~\mathcal{H}_{-1}(x,t)=0.
\end{array} }\right.
\end{eqnarray}
 For initial condition $t=0$, we have:
\bal\label{2}
\mathcal{H}_{n+1}(x,0)=\dfrac{x}{n+1}\mathcal{H}_{n}(x,0)~~~~or ~~~\mathcal{H}_{n}(x,0)=\dfrac{x^n}{n!}.
\eal
Therefore, considering \eqref{2}, for $t\in [0,T]$, it is concluded that
\bal\label{Hh}
\mathcal{H}_{n}(x,t)=\frac{1}{n!}\bar{H}_{n}(x,t)=\cas{\frac{\sqrt{t^n}}{n!}.H_{n}(\frac{x}{\sqrt{t}})~~&t\neq 0,\\ \dfrac{x^n}{n!}~~~~&t=0.}
\eal
 Thus, using recursion formula \eqref{recursive2}, some of these Hermite polynomials compute as follows:
\bals
\mathcal{H}_{0}(x,t)=1~~~,~~~\mathcal{H}_{1}(x,t)=x~~~,~~~ \mathcal{H}_{2}(x,t)=\frac{x^2}{2}-\frac{t}{2},\\
\mathcal{H}_{3}(x,t)=\frac{x^3}{6}-\frac{tx}{2}~~~,~~~ \mathcal{H}_{4}(x,t)=\frac{x^4}{24}-\frac{tx^2}{4}+\frac{t^2}{8}.
\eals
 Since the classical Hermite polynomials of standard Gaussian random variables $N(0,1)$ are orthogonal, we have:
\bal
\frac{\delta_{mn}}{n!}=&\int_{-\infty}^{\infty}H_{n}(x)H_{m}(x)\frac{1}{\sqrt{2\pi}}e^\frac{-x^2}{2}dx
=\int_{-\infty}^{\infty}H_{n}(\frac{x}{\sqrt{t}})H_{m}(\frac{x}{\sqrt{t}})\frac{1}{\sqrt{2\pi t}}e^\frac{-x^2}{2t}dx\\\notag
&=\int_{-\infty}^{\infty}\bar{H}_{n}(x,t)\bar{H}_{m}(x,t)\frac{1}{\sqrt{t^{n+m}}}\gamma_{t}dx
=\int_{-\infty}^{\infty}\mathcal{H}_{n}(x,t)\mathcal{H}_{m}(x,t)\frac{{n!m!}}{\sqrt{t^{n+m}}}\gamma_{t}dx.
\eal
 Thus, the sequence $\{\mathcal{H}_{n}(x,t)\}_{n=0}^{\infty}$ is an orthogonal set in $\LL ^{2}_{\gamma_t}(\RR\times[0,T])$:
\bal\label{Horthogonal}
\<\mathcal{H}_{n},\mathcal{H}_{m}\>_{\gamma_{t}}=\int_{0}^{T}\int_{-\infty}^{\infty}
\mathcal{H}_{n}(x,t)\mathcal{H}_{m}(x,t)\gamma_{t}dxdt=\frac{T^{n+1}}{n+1!}\delta_{mn}.
\eal
Similar to the other orthogonal polynomials, 2-D Hermite polynomials enjoy a generating function \cite{Evance}:
\begin{eqnarray}\label{martingale}
Z_{t}^{\lambda}=e^{\lambda x-\frac{\lambda^2}{2}t}=\sum_{n=0}^{\infty}\lambda^{n}.\mathcal{H}_{n}(x,t),~~~~\lambda\in\RR.
\end{eqnarray}
The generating function \eqref{martingale} is a crucial tool for studying the stochastic Hermite polynomials properties. We will utilize it frequently in our later derivations and results. If we replace $x$ with the Wiener process $W (t) $, this stochastic process is named \emph{Exponential Martingale} \cite{Evance} \cite{Oksendal}. Applying $It\hat{o}$'s formula on this process, we obtain the following SDE:
\bal\label{Z}
\cas{dZ_{t}^{\lambda}=\lambda.Z_{t}^{\lambda} dW_{t}, \\
Z_{t}^{\lambda}(0)=1.}
\eal
 Substituting $\sum_{n=0}^{\infty}\lambda^{n} \mathcal{H}_{n}(x,t)$ for the Exponential Martingale $Z_{t}^{\lambda}$ in \eqref{Z}, we can infer the following theorem \cite{Evance} that is well-known as the stochastic calculus with the stochastic Hermite polynomials.\\
\begin{theorem}
 Let $W_{t}$ be a Wiener process and $\mathcal{H}_{n}(W_{t},t)$ be the $n-th$ order stochastic Hermite polynomial in terms of $W_{t}$. For each $t\geq0$ and $n=0,1,2,\cdots$, we have:
 \bal\label{dhn}
 d\mathcal{H}_{n+1}(W_{t},t)=\mathcal{H}_{n}(W_{t},t)dW_{t}.
 \eal
\end{theorem}
 This theorem implies that the stochastic Hermite polynomials $\mathcal{H}_{n}(W_{t},t)$ with respect to the filtration $\{\mathcal{F}_{t}\}$ and measure $\rho$ are Martingale processes.\\
 According to \cite{Shen,G}, for a given positive weight function $\omega_{n}(t)=t^n$ defined in time interval $(0,T)$, there
exists a unique family of monic Orthogonal polynomials $~\{\mathcal{P}_{m,n}(t)\}_{m=0}^{\infty}$ as basis of $\LL ^{2}_{\omega_{n}}$ generated by a three-term recursion relationship such that:
\begin{eqnarray}\label{Gmn}
\int_{t=0}^{T}t^{n}\mathcal{P}_{r,n}(t)\mathcal{P}_{s,n}(t)dt=\delta_{rs}.
\end{eqnarray}
 The sequence $~\{\mathcal{P}_{m,n}(t)\mathcal{H}_{n}(x,t)\}_{m,n=0}^{\infty}$ is also an orthogonal set on the space $\LL ^{2}_{\gamma_t}(\RR\times[0,T]).$ For each $m,n\in \NN \cup \{0\}$, we possess:
\bals
\<\mathcal{P}_{r,m}\mathcal{H}_{m},\mathcal{P}_{s,n}\mathcal{H}_{n}\>_{\gamma_{t}}=&\int_{0}^{T}\int_{-\infty}^{\infty}\big(\mathcal{P}_{r,m}(t)\mathcal{H}_{m}(x,t)\big)\big(\mathcal{P}_{s,n}(t)\mathcal{H}_{n}(x,t)\big)\gamma_t dx dt=\\
&\int_{0}^{T}(\mathcal{P}_{r,m}(t)\mathcal{P}_{s,n}(t))\big(\int_{-\infty}^{\infty}\mathcal{H}_{m}(x,t)\mathcal{H}_{n}(x,t)\gamma_t dx\big) dt=\\
&\int_{0}^{T}(\mathcal{P}_{r,m}(t)\mathcal{P}_{s,n}(t))\big(\frac{t^{n}}{n!}\delta_{mn}\big)dt=\frac{\delta_{mn}\delta_{rs}}{n!}.
\eals
 Since the linear span of polynomials $\{t^{n} x^{m}\}$ for $m,n\in\NN\cup\{0\}$ are dense in $\LL^{2}(\RR\times[0,T] ,\rho)$, in order to indicate that $~\{ \mathcal{P}_{m,n}(t)\mathcal{H}_{n}(x,t)\}_{m,n=0}^{\infty}$ is also an orthogonal basis, it would be enough to demonstrate all these polynomials can be generated by them.
 \begin{theorem}
  Each polynomial in terms of variables $t$ and $x$ can be generated by the orthogonal set $~\{ \mathcal{P}_{m,n}(t)\mathcal{H}_{n}(x,t)\}_{m,n=0}^{\infty}$.
 \end{theorem}
\textbf{ proof}: For each polynomial set $\{t^{n} x^{m}\}_{n=0}^{\infty}$, proof is done by induction on $m\in\NN\cup\{0\}$. For $m=0$, since $t^{n} \in{span}\{ \mathcal{P}_{i,0}(t)\}_{i=0}^{\infty}$, we have:
\bals
t^{n}=\sum_{i=0}^{\infty}a_{i,0}\mathcal{P}_{i,0}(t)\mathcal{H}_{0}(x,t).
\eals
 Similarly, for $m=1$, since $t^{n}\in span\{ \mathcal{P}_{i,1}(t)\}_{i=0}^{\infty}$, we have:
\bals
t^{n} x =\sum_{i=0}^{\infty}a_{i,1}\mathcal{P}_{i,1}\mathcal{H}_{1}(x,t).
\eals
Assume for $m$ we have:
\bals
t^{n} x^{m} =\sum_{i=0}^{\infty}\sum_{k=0}^{m}a_{i,m}\mathcal{P}_{i,k}(t)\mathcal{H}_{k}(x,t).
\eals
We must show that the same formula is true for $(m+1)$:
\bals
t^{n} x^{m+1} = t^{n} x^{m} x = \big(\sum_{i=0}^{\infty}\sum_{k=0}^{m}a_{i,k}\mathcal{P}_{i,k}(t)\mathcal{H}_{k}(x,t)\big)\mathcal{H}_{1}(x,t).
\eals
 On the other hand, using recursive equality \eqref{recursive2}, we get:
\bals
(k)\mathcal{H}_{1}(x,t)\mathcal{H}_{k}(x,t) = (k+1) \mathcal{H}_{k+1}(x,t)+t \mathcal{H}_{k-1}(x,t).
\eals
Afterward, we have:
\bals
t^{n} x^{m+1} = \sum_{i=0}^{\infty}\sum_{k=0}^{m}\big((k+1)a_{i,k}\mathcal{P}_{i,k}(t)\big)\mathcal{H}_{k+1}(x,t)+
\sum_{i=0}^{\infty}\sum_{k=0}^{m}\big(t a_{i,k}\mathcal{P}_{i,k}(t)\big)\mathcal{H}_{k-1}(x,t).
\eals
 But since $(k+1)a_{i,k}\mathcal{P}_{i,k}(t)\in Span\{ \mathcal{P}_{i,k+1}\}_{i=0}^{\infty}$ and $t a_{i,k}\mathcal{P}_{i,k}(t)\in Span\{ \mathcal{P}_{i,k-1}(t)\}_{i=0}^{\infty}$, the following equality is obtained:
\bals
t^{n} x^{m+1} =\sum_{i=0}^{\infty}\sum_{k=0}^{m+1}b_{i,k}\mathcal{P}_{i,k}(t)\mathcal{H}_{k}(x,t),
\eals
 and consequently the proof is completed by induction.~~~~~~~~~~~~~~~~~~~~~~~~~~~~~~~~~~~~~~~~~~~~~~~~~~~~~~~$\Box$\\

 Because $~\{ \mathcal{P}_{m,n}(t)\mathcal{H}_{n}(x,t)\}_{m,n=0}^{\infty}$ is an orthogonal basis in $\LL ^{2}_{\gamma_t}(\RR\times[0,T])$, therefore for any $f$ that  belongs to this space we have :
\begin{eqnarray}\label{WCE}
f(x,t)=\sum_{n=0}^{\infty}\sum_{m=0}^{\infty}a_{m,n}\mathcal{P}_{m,n}(t)\mathcal{H}_{n}(x,t)~~~
\end{eqnarray}
where 
\begin{equation}
a_{m,n}=n!\<f,\mathcal{P}_{m,n}\mathcal{H}_{n}\>_{\gamma_{t}}.
\end{equation}
  Defining $f_{n}(t)=\sum_{m=0}^{\infty} a_{m,n}\mathcal{P}_{m,n}(t)$ and applying \eqref{Horthogonal}, we obtain:
\begin{eqnarray}\label{WCE}
f(x,t)=\sum_{n=0}^{\infty}f_{n}(t).\mathcal{H}_{n}(x,t)~~~~~~,~~~~~~~f_{n}(t)=\dfrac{n!}{t^n}\int_{-\infty}^{\infty}
f(x,t)\mathcal{H}_{n}(x,t)\gamma_{t}dx
\end{eqnarray}

\section{An expansion method for solving SDEs }
In this section, we consider the general form of the stochastic differential equation:
\bal\label{SDEs}
\cas{dX_{t}=F(X_{t},t)dt+G(X_{t},t)dW_{t},\\X(0)=X_0, }
\eal
 where the drift and the diffusion coefficients $F$ and $G$ belong to $\LL^1(\RR\times[0,T]) $ and $ \LL^2(\RR\times[0,T])$, respectively and we assume that relation \eqref{pa11} holds therefore the equation \eqref{SDEs} has a unique path-independent solution.
  According to the \eqref{WCE}, stochastic process $X_{t}=X(W_t,t)$ yields in the following expansion:
\bal\label{XFG}
X_{t}=X(W_{t},t)=\sum_{i=0}^{\infty}a_{i}(t)\mathcal{H}_{i}(W_{t},t)~~~
\eal
Because $W_t$ is a normal random variable with mean zero and variance $t$, we could obtain $a_i(t)$ as below:
\begin{equation}
a_{i}(t)=\dfrac{i!}{t^i}\int_{-\infty}^{\infty}X_t\mathcal{H}_{i}(x,t)\gamma_{t}dx=E[X_{t}.\mathcal{H}_{i}(W_t,t)].
\end{equation}

Consequently, the definition of $a_{n}(t)$  and \emph{parseval's equality} lead to calculating the expectation and variance of the stochastic process $X(W_t,t)$ in each point of time interval $[0,T]$ by corresponding expansion coefficients:
\bal\label{EVar}
E[X(W_t,t)]=a_{0}(t)~~,~~E[X^{2}(W_t,t)]=\sum_{n=0}^{\infty}\dfrac{t^n}{n!}a_{n}^{2}(t).
\eal

 For each two It$\hat{o}$ process $X_{t},Y_{t}\in \LL ^{2}_{\gamma_t}(\RR\times[0,T])$ , we have the following equality named It$\hat{o}$ product formula: 
\bals
 d(X_{t}Y_{t})=Y_{t}dX_{t}+X_{t}dY_{t}+dX_{t}dY_{t}.
\eals
 Now as the first step to find the unknown coefficients $a_{i}(t)$, we apply It$\hat{o}$ product formula for equations\eqref{SDEs} and \eqref{dhn}.
\bal
&d(X_{t}\mathcal{H}_{n}(W_{t},t))=X_{t}d\mathcal{H}_{n}(W_{t},t)+\mathcal{H}_{n}(W_{t},t)dX_{t}+dX_{t}d\mathcal{H}_{n}(W_{t},t)\\\notag
&=\Big(\mathcal{H}_{n}(W_{t},t)F(X_{t},t)+\mathcal{H}_{n-1}(W_{t},t)G(X_{t},t)\Big)dt+\Big(\mathcal{H}_{n}(W_{t},t)G(X_{t},t)+X_{t}\mathcal{H}_{n-1}(W_{t},t)\Big)dW_{t}.
\eal
Rewriting this equation in integral form and taking expectation, we infer the following equation:
\bal
E_{}[X_{t}\mathcal{H}_{n}(W_{t},t)]=X_{0}\delta_{0n}+E_{}\Big(\int_{0}^t[\mathcal{H}_{n}(W_{s},s)F(X_{s},s)]+[\mathcal{H}_{n-1}(W_{s},s)G(X_{s},s)]ds\Big).
\eal
  As a matter of fact, since $\int_{0}^t (\mathcal{H}_{n}(W_{s},s)G(X_{s},s)+X_{s}\mathcal{H}_{n-1}(W_{s},s))dW_{s}$ is an $It\hat{o}$ integral and also a martingale process, its expectation is equal to zero. Afterward, employing the Fubini's theorem which allows the order of integration to be changed in iterated integrals we have;
\bal
a_{n}(t)\dfrac{t^{n}}{n!}=X_{0}\delta_{0n}+\int_{0}^t\Big(E_{}[\mathcal{H}_{n}(W_{s},s)F(X_{s},s)]+E_{}[\mathcal{H}_{n-1}(W_{s},s)G(X_{s},s)]\Big)ds
\eal
 Thereby, considering Eq.\eqref{XFG} and taking derivation, we reach a system of integro-differential equations:
\bal\label{F}
a_{n}(t)\dfrac{t^{(n-1)}}{(n-1)!}+a_{n}^{'}(t)\dfrac{t^{n}}{n!}=\Big(E_{}[\mathcal{H}_{n}(W_{t},t)F(X_{t},t)]+E_{}[\mathcal{H}_{n-1}(W_{t},t)G(X_{t},t)]\Big)
\eal
 and so:
\bal\label{Final}
a_{n}^{'}(t) = \dfrac{n!}{t^n}\Big(E[\mathcal{H}_{n}(W_t,t)F(X(W_t,t),t)]+E[\mathcal{H}_{n-1}(W_t,t)G(X(W_t,t),t)]\Big)-\frac{n}{t}a_{n}(t)
\eal
  For solving this system of equations, we need to extract coefficients $a_{i}(0)$,($i=0,1,2,...$) as initial conditions. For $n=0$, we get $a_{0}(t)=X_0+\int_0^t\Big(E[F(X(W_s,s),s)]\Big)ds$, and therefore $a_{0}(0)=X_0$. For $n\neq 0$, since $a_{n}(t)$ is continuous we have:
 \bals
 a_{n}(0)= lim_{t\rightarrow 0}\dfrac{n!}{t^n}\big(\int_0^t\Big(E[\mathcal{H}_n(W_s,s)F(X(x,s),s)]+E[\mathcal{H}_{n-1}(W_s,s)G(X(W_s,s),s)]\Big)ds\big)
 \eals
  Applying \emph{L'Hopital's Rule} into computing function limits, $a_{n}(0)$ can be calculated as below:
 \bal\label{lopital}
 a_{n}(0)=lim_{t\rightarrow 0}\dfrac{(n-1)!}{t^{n-1}}\big(E[\mathcal{H}_n(W_t,t)F(\hat{X}(W_t,t),t)]+E[\mathcal{H}_{n-1}(W_t,t)G(\hat{X}(W_t,t),t)]\big).
 \eal
  So that according to the Euler approximation method $\hat{X}(W_t,t)=X_{0}+F(X_{0},0)t+G(X_{0},0)W_t$ has been put in place of $X_{t}$.
 Afterwards, we consider a finite dimensional case of this system of integro-differential equations to compute truncation form of $X_{t}$ as follows:
\bal
X_{N}(W_{t},t)=\sum_{n=0}^{N}a_{n}(t)\mathcal{H}_{n}(W_{t},t).
\eal
 After discretization of the time interval with $t_{i}= ih~~(i=0,1,2,...,M)$, where $h=T/M$, the subsequence equality is achieved as
\bals
a_{n}^{'}(t_{i}) = \dfrac{n!}{t_{i}^n}\Big(E[\mathcal{H}_{n}(W_{t_i},t_{i})F(X_{N}(W_{t_i},t_{i}),t_{i})]+E[\mathcal{H}_{n-1}(W_{t_i},t_{i})G(X_{N}(W_{t_i},t_{i}),t_{i})]\Big)-\frac{n}{t_{i}}a_{n}(t_{i}).
\eals
 But due to the singularity of the right side of this system of integro-differential equations in $t_{0}=0$, we can not perform Rung-Kutta method from $t_{1}= h$. To find a significant approximation of $a_{n}(h)$ and in order to overcome the singularity in $t=0$, we use the Euler method. Substituting $(a_{n}(h)-a_{n}(0))/h$ for $a_{n}^{'}(h)$, the preceding equation converts to a system of integro-differential equations:
\bals
a_{n}(h) =\frac{1}{n+1} \big(\dfrac{n!}{h^{(n-1)}}\Big(E[\mathcal{H}_n(W_{h},h)F(X(W_{h},t_{h}),h)]+E[\mathcal{H}_{n-1}(W_{h},h)G(X(x,t_{h}),h)]\Big)+a_{n}(0)\big).
\eals
 The numerical solution of final system\eqref{Final}, can be done with various one step methods as Euler and Rung- kutta of different orders\cite{Hoog}. In this paper, the numerical solution is computed based on second-order Rung-Kutta method.
 
 \section{Analytical and numerical results}
\textbf{Example 1}
Consider the {\textbf{Geometric Brownian Motion}} (GBM) model
\bal
\cas{dX_{t}=\mu(t) X_{t}dt+\sigma X_{t}dW_{t},\\X(0)=X_{0}.}
\eal
where $\mu(t)$ and $\sigma$ are the time dependent and constant parameter, respectively. From the system of equations \eqref{Final}, we obtain the following equations:
\bal
a_{n}^{'}(t)&= \dfrac{n!}{t^n}\Big(E[\mathcal{H}_{n}(W_{t},t)\mu(t)X(W_{t},t)]+E[\mathcal{H}_{n-1}(W_{t},t)\sigma X(W_{t},t)]\Big)-\frac{n}{t}a_{n}(t)\\\notag
&=\mu(t)a_{n}(t)+\sigma\frac{n}{t}a_{n-1}(t)-\frac{n}{t}a_{n}(t)= \big(\mu(t)-\frac{n}{t}\big)a_{n}(t)+\sigma\frac{n}{t}a_{n-1}(t).
\eal
It can be verified that $a_{n}(t)=\sigma^n e^{\int_0^t\mu(s) ds }$ is the explicit solution of this system. Therefore,
\bal
X_{t}&=\sum_{n=0}^{\infty}\sigma^ne^{\int_0^t\mu(s)ds} \mathcal{H}_{n}(W_{t},t)=e^{\int_0^t\mu(s)ds}\sum_{n=0}^{\infty}\sigma^n\mathcal{H}_{n}(W_{t},t).
\eal
 From \eqref{martingale}, we have $\sum_{n=0}^{\infty}\sigma^n\mathcal{H}_{n}(W_{t},t) = \exp(\int_0^t\sigma dW_{s}-\int_0^t\frac{\sigma^2}{2}ds)$, and finally the exact solution of the stock pricing model equation is:
  $$X_{t}= \exp\Big(\int_0^t\sigma dW_{s}+(\int_0^t\mu(s)-\frac{\sigma^2}{2})ds\Big).$$
 In a particular case $\mu (t)=0$ the obtained solution is the exponential martingale process.\\
  \begin{table}
\caption{Numerical results of Example 1 for different values of N and T.}
\label{tab:1}       
\begin{tabular}{llllllll}
\hline\noalign{\smallskip}
\quad&\quad&\quad&\quad N=30 &\quad&\quad &\quad N=40 &\quad \\
& \quad T &\quad\quad EM &\quad\quad PM &\quad\quad Exact &\quad\quad EM &\quad\quad PM&\quad\quad Exact \\
\noalign{\smallskip}\hline\noalign{\smallskip}
& 3   & \quad 8.276397  &  \quad 4.575074 &  \quad 4.301673 &  \quad 13.521179 &\quad 7.813891 &\quad 7.848716\\
&2    & \quad 1.410893  &  \quad 0.832558 &  \quad 0.855598 &  \quad 2.129019  &\quad  1.35756 &\quad 1.363187\\
& 1.5 & \quad 1.850382  &  \quad 1.201318 &  \quad 1.192680 &  \quad1.0618345  &\quad 0.762551 &\quad 0.765961\\
&1    & \quad0.5971905  &  \quad 0.462382 &  \quad 0.465634 &  \quad0.8095762  &\quad 0.637728 &\quad 0.638863\\
\hline
\noalign{\smallskip}\hline
\end{tabular}
\end{table}
\begin{figure}
\centering
\begin{tabular}{cc}
\includegraphics[scale=0.27]{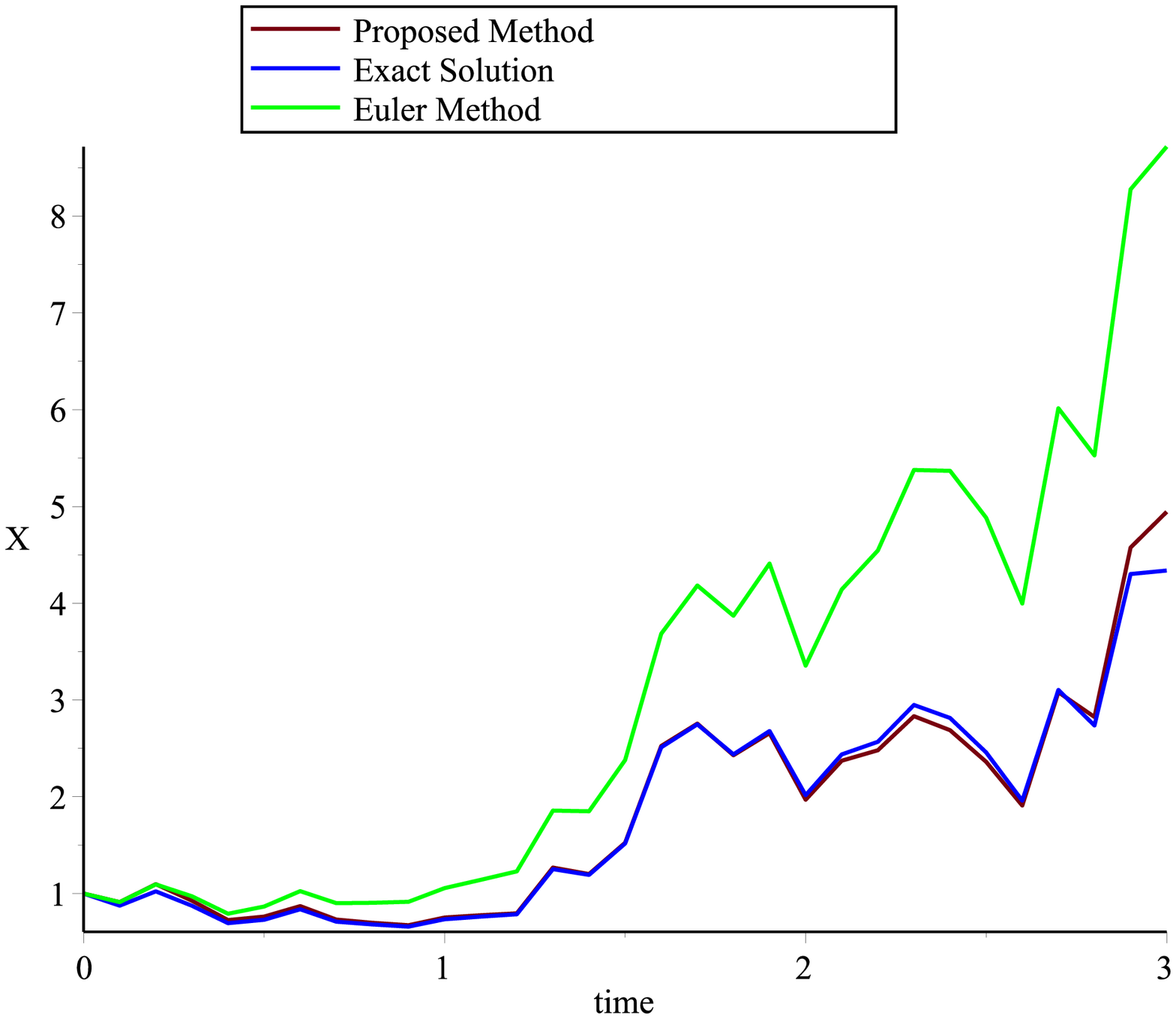} &\includegraphics[scale=0.27]{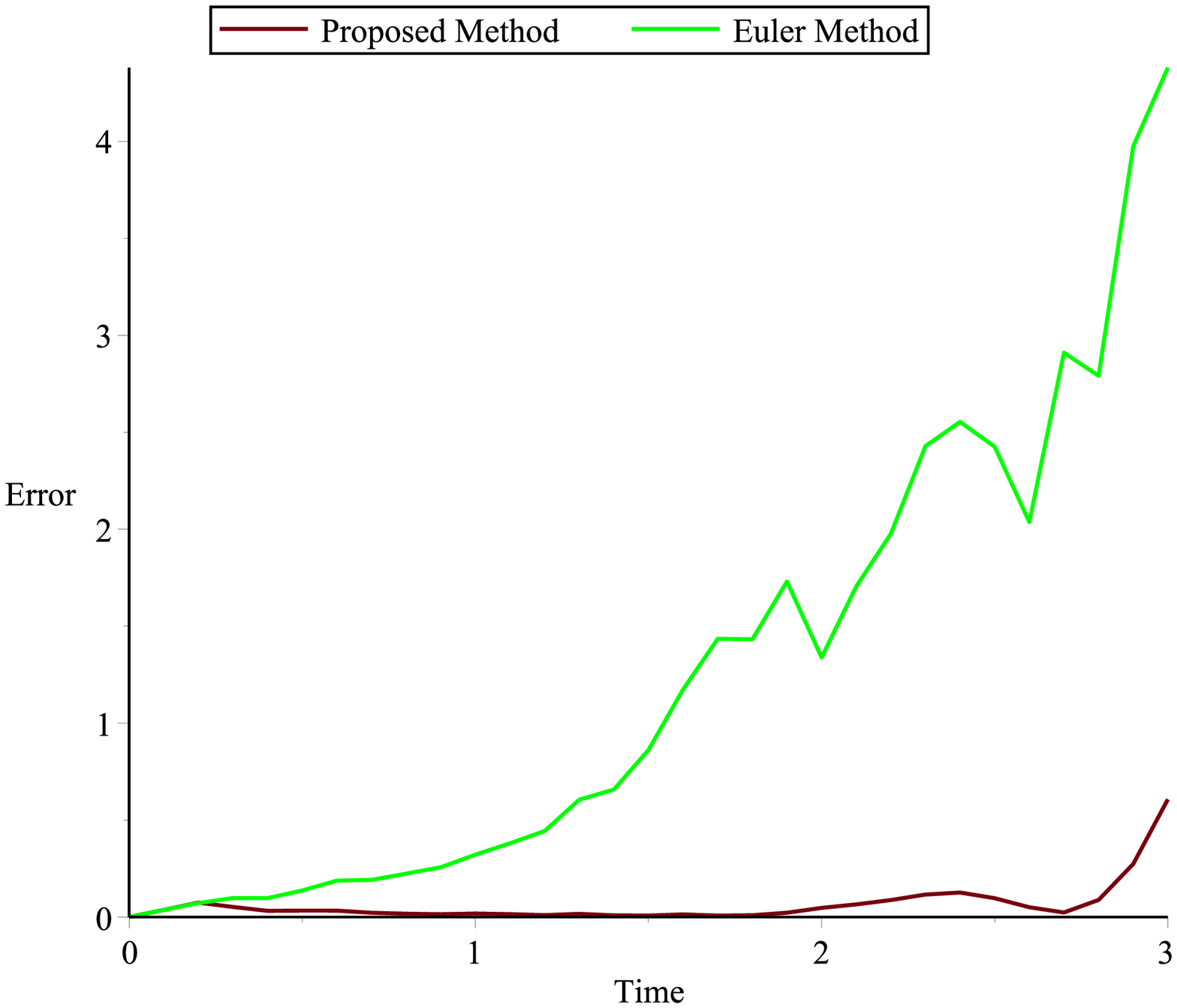}\\
a&b
\end{tabular}
\caption{a) The graphs of the approximate and the exact solutions of Example 1 for maturity time $T=3$, $N=4$ and number of points $M=30$ and b) The graphs of absolute error.}
\end{figure}

\textbf{Example 2} Consider \emph{\textbf{Cox-Ingersoll-Ross investment Model}} in the following special case
\bal\label{cox}
\cas{dX_{t}=(\frac{K\sigma '(t)}{\sigma(t)} X_{t}+\sigma^{2K}(t))dt+2\sigma ^K(t) \sqrt{X_{t}}dW_{t},\\X(0)=X_{0},}
\eal
where $\sigma(t)>0$ and $\sigma(0)=1$. By applying It$\hat{o}$ formula on $U_{t}=\sqrt{X_{t}}$, we reach to the equation the prominent stochastic model Ornschten-Uhlenberg (Langevin process):
 \bal\label{langevin}
\cas{dU_t=\frac{K\sigma '(t)}{\sigma(t)}U_t dt+\sigma ^K(t) dW_{t},\\U_0=\sqrt{X_{0}}}.
\eal
The exact solution of this equation by applying It$\hat{o}$ formula on $Ue^{-\int_{0}^{t}\frac{K\sigma '(s)}{\sigma(s)}ds}=U \sigma ^{-K}(t)$ is:

\bal
U_{exact}=\sqrt{X_{t}}=e^{\int_{0}^{t}\frac{K\sigma '(s)}{\sigma(s)}ds}(\sqrt{X_{0}}+\int_{0}^{t}e^{\int_{0}^{s}-\frac{K\sigma '(r)}{\sigma(r)}dr}\sigma ^K(s)dW_{s})=\sigma ^K(t)(\sqrt{X_{0}}+W_{t}).
\eal

 In order to solve the equation \eqref{langevin} by Galerkin method based on 2-D Hermite polynomials, we put $U_t=\sum_{n=0}^{\infty}a_{n}(t)\mathcal{H}_{n}(W_{t},t)$. Now multiplying Eq.\eqref{langevin} by $\mathcal{H}_{n}(W_{t},t)$, and taking expectation and derivation respectively, we obtain the following O.D.E. similar to \eqref{F}:
\bal
E'[U_t \mathcal{H}_{n}(W_t,t)]=\frac{K\sigma '(t)}{\sigma(t)} E[U_t \mathcal{H}_{n}(W_t,t)]+\sigma ^K(t) E[\mathcal{H}_{n-1}(W_t,t)]
\eal
\begin{figure}
\centering
\begin{tabular}{cc}
\includegraphics[scale=0.25]{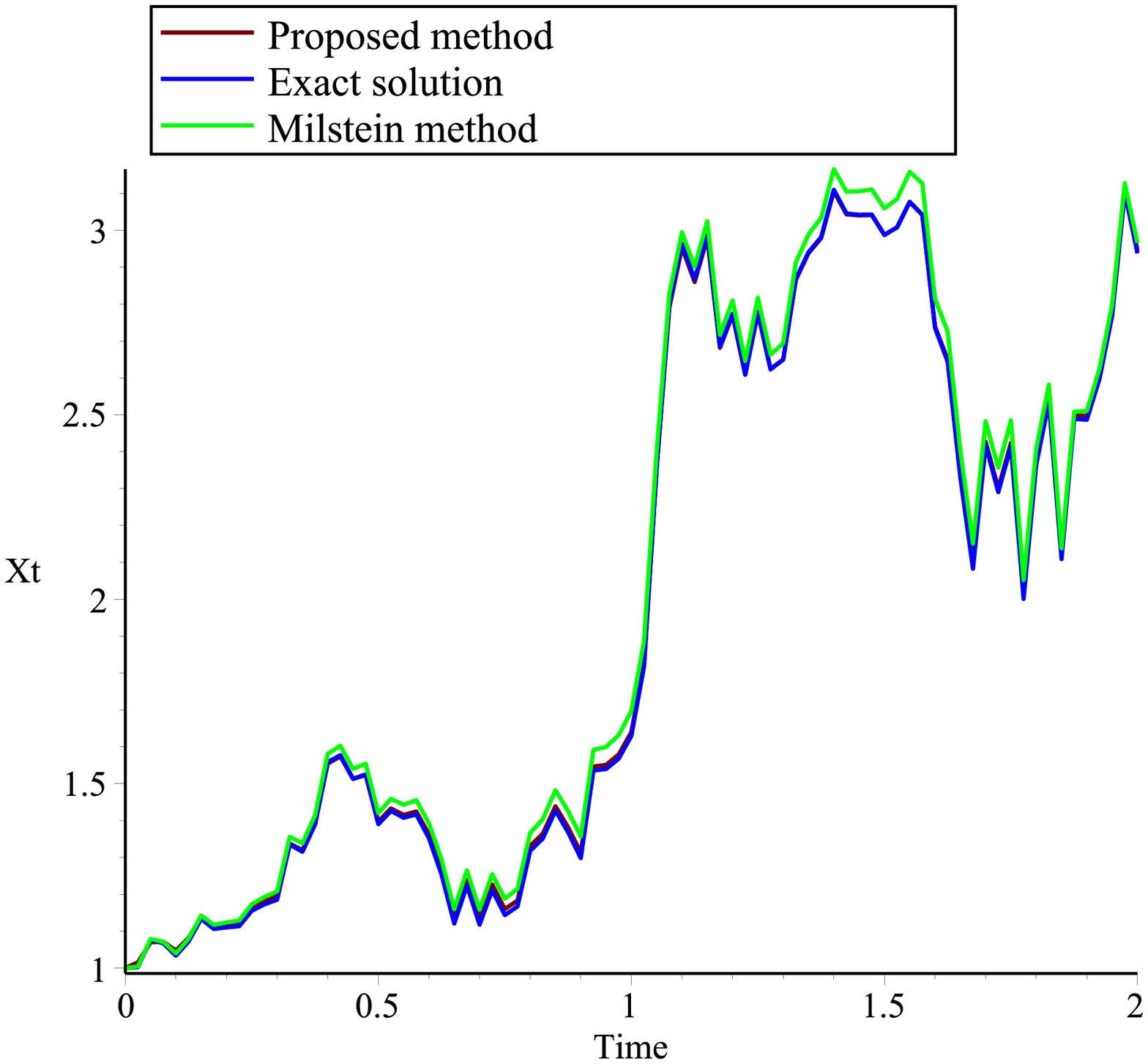} &\includegraphics[scale=0.25]{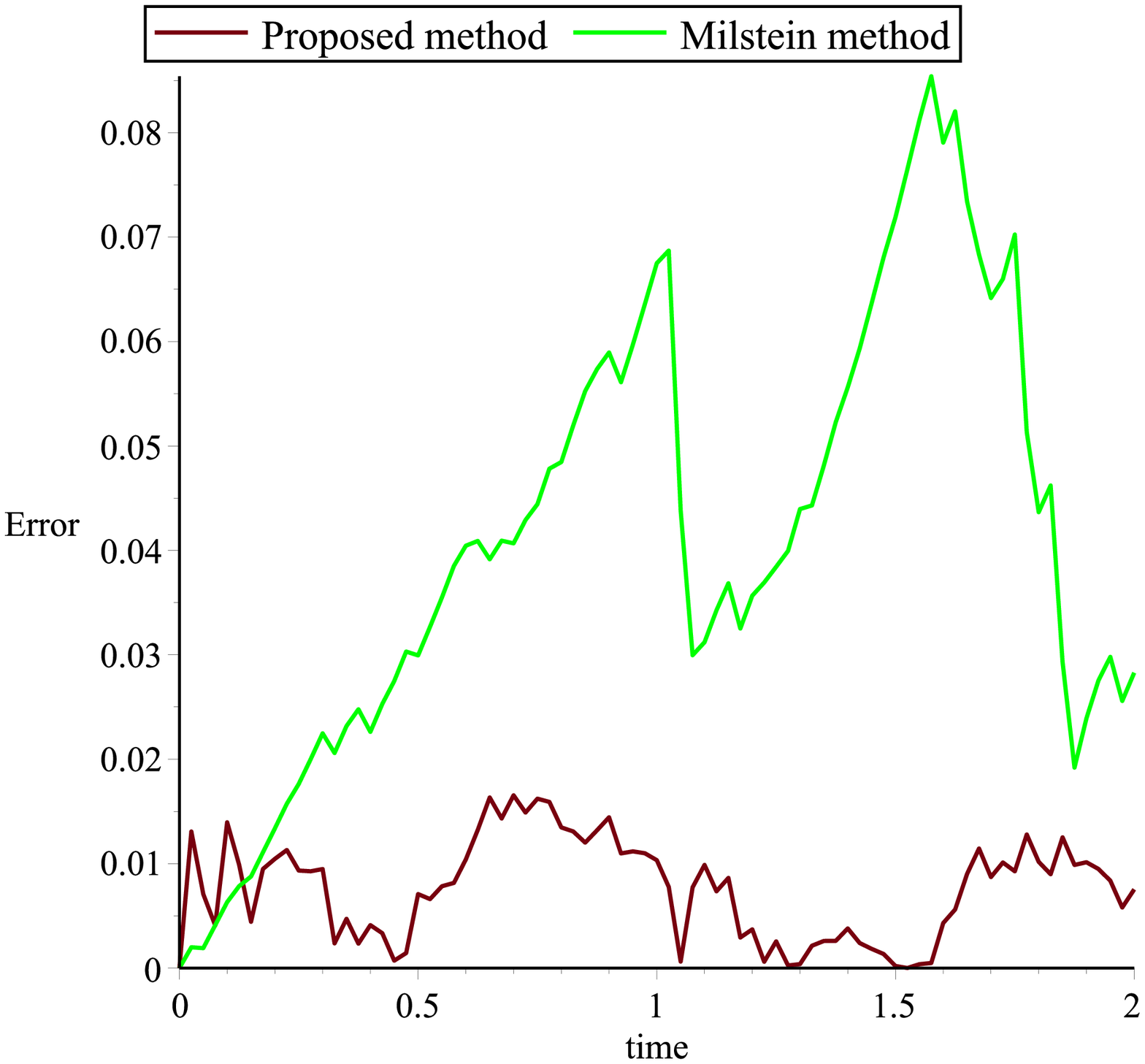}\\
a&b
\end{tabular}
\caption{a) The graphs of approximate and exact solution of Cox-Ingersoll-Ross model $dX_{t}=(X_{t}/(t+1)+(t+1)/16)dt+(\sqrt{(t+1)X_{t}}/2)dW$, with initial condition $X(0)=1$, for $T=2$,$N=5$  and number of points $ M=80.$
 b) The graphs of absolute error.  }
\end{figure}
  Considering different initial conditions for indices $n$, we could find $a_{n}(t)$ by solving some equations in cases $n=0$, $n=1$ and $n\geq 2$, respectively:
\bals
\cas{E'[U_t]=\frac{K\sigma '(t)}{\sigma(t)} E[U_t],\\E[U_0]=U(0)=\sqrt{X_{0}}.}
\eals
\bals
\cas{E'[U_t\mathcal{H}_{1}(W_t,t)]=\frac{K\sigma '(t)}{\sigma(t)} E[U_t\mathcal{H}_{1}(W_t,t)]+\sigma ^K(t),\\E[U_t\mathcal{H}_{1}(W_t,t)](0)=a_{1}(0)=1.}
\eals
\bals
\cas{E'[U_t\mathcal{H}_{n}(W_t,t)]=\frac{K\sigma '(t)}{\sigma(t)} E[U_t\mathcal{H}_{n}(W_t,t)],\\E[U_t\mathcal{H}_{n}(W_t,t)](0)=a_{n}(0)=0.~~(n\geq 2)}
\eals
Consequently, answers of these equations are
 \bals
 &E[U_t]= a_{0}(t)=\sqrt{X_{0}}e^{\int_{0}^{t}\frac{K\sigma '(s)}{\sigma(s)}ds}=\sqrt{X_{0}}\sigma ^K(t).\\
 &E[U_t\mathcal{H}_{1}(W_t,t)]=t a_{1}(t)=e^{\int_{0}^{t}\frac{K\sigma '(s)}{\sigma(s)}ds}(\int_{0}^{t}e^{\int_{0}^{s}-\frac{K\sigma '(r)}{\sigma(r)}dr}\sigma ^K(s)ds)=t\sigma ^K(t).\\
 &E[U_t\mathcal{H}_{n}(W_t,t)]=\dfrac{t^n}{n!} a_{n}(t)=E[U_t\mathcal{H}_{n}(W_t,t)]=0.
\eals

Finally, we get the following equation:
  \bal
 U_t=\sqrt{X_{t}}=\sigma ^K(t)(\sqrt{X_{0}}+W_{t}).
 \eal
 Figure 2 depicts a different numerical example of this model with stochastic differential equation
 \bal
 dX_{t}=(\frac{X_{t}}{t+1}+\frac{t+1}{16})dt+(\frac{\sqrt{(t+1)X_{t}}}{2})dW,
 \eal
  with initial value $X(0)=1$ and also absolute error for $T=2$ with $M=80$ points.

\textbf{Example 3}
 Consider non-linear time dependent stochastic differential equation
\bal
\cas{dX_{t}=\big(-a^2 sinX_{t} cos^3X_{t}\big) dt+a cos^2X_{t}dW_{t},~~(a\in\RR)\\X(0)=X_{0}.}
\eal
 The exact solution of this equation is $X_{t}=arctan(aW_{t}+tanX_{0}).$ The numerical solution of this equation based on proposed method with $N=5$ and $M=100$ is computed and the result is compared with $\emph{Milstein}'$ method in Figure 3.  
\begin{figure}
\centering
\begin{tabular}{cc}
\includegraphics[scale=0.25]{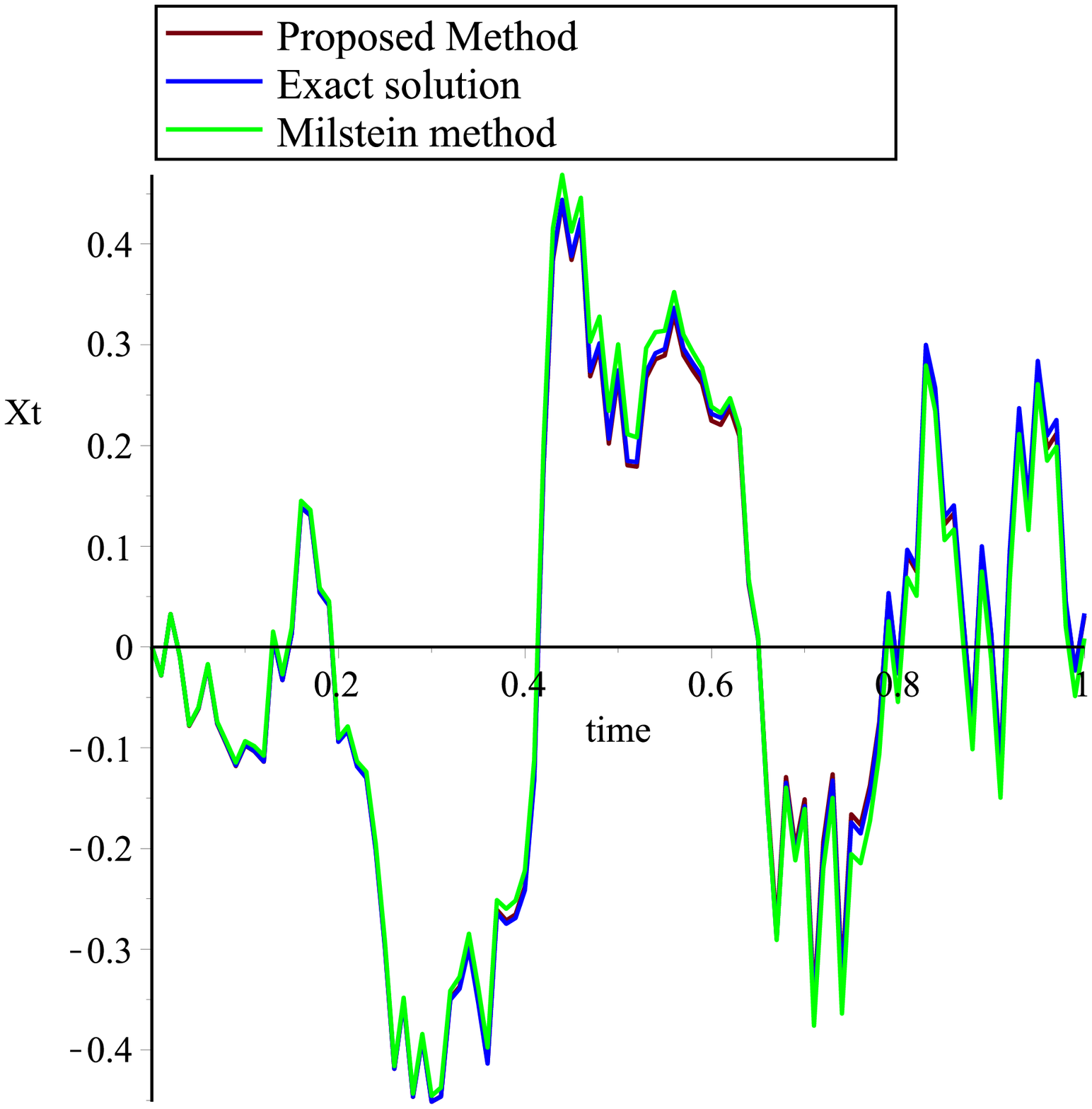} &\includegraphics[scale=0.25]{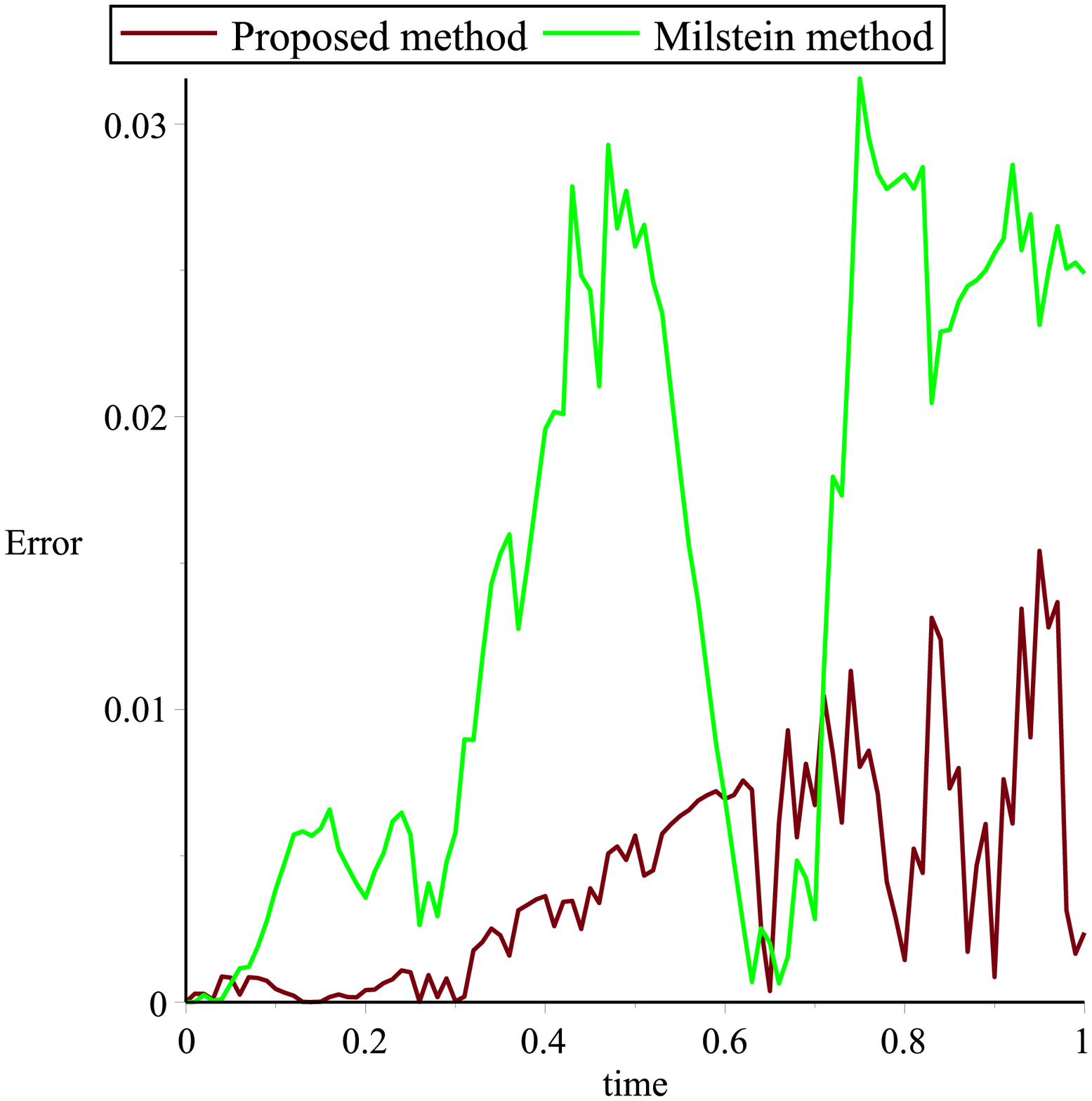}\\
a&b
\end{tabular}
\caption{a) The graphs of approximate and exact solution of Example $3$ for $T=1$, $N=5$, and number of points $ M=100$ and b) Figure of absolute error. }
\end{figure}

\section{ conclusion}
In this article, we introduced an orthogonal basis expansion method to solve stochastic differential equations with a path\-independent solution.
 For a truncated form of the solution, the equation reached a closed nonlinear system of deterministic integro-differential equations for the related coefficients.
 The orthogonal basis expansion method provided careful analytical formulas for computing statistical moments such as mean and variance, which in this paper even the statistical moments up to the fourth order, was found. In the numerical experiments with stochastic equations, we compared the
solution accuracy and the computational effectiveness of the orthogonal basis expansion and other stochastic numerical methods like the Monte Carlo (MC), E.M. and Milstein method.


\end{document}